# Principle "synthesis" for the solution of tasks of class NP

Rustem Valeyev, Saint Petersburg

*C o n t e n t s*

## Subject

Mathematical experts couldn't find any algorithm of obtaining the exact answer for one mass NP-complete task during four decades after emergence of the concept "class NP". It is logical to assume that perfectly developed traditional and/or developed experimentally approaches to creation of algorithms and will not provide considerable progress in this question. Therefore, despite of the fact what the final answer to "P ? NP" dilemma will be, it is essential now to look for other ideologies of algorithms functioning of the solution of similar tasks on discrete structures. "Metaedroalgorithm" can be one of such nonconventional, but very perspective approaches.

## 1. Basic metaedroalgorithm

The meaning of the offered term "metaedroalgorithm" (or, as a variant, "metaalgorithm") and ultimate goal of such set of operations over initial data of a solved task are extremely simple: it is not calculation, not detection of its exact answer existing somewhere, and independent creation (synthesis) of the required answer. The principle considered below in the most general terms "synthesis" is reduced to selection of such set of fragments which is closer to objectively existing exact answer of a solved individual task on discrete structures.

The prefix "meta" here means the principle of algorithm's creation on the basis of "all-information" approach to business, and that it treats the algorithms suitable for many mass tasks at the same time. "Edr" specifies full analogy of creation of intangible virtual (information) object to creation of a material physical body. "Basic" specifies only initial main idea, only the beginning of "trunk" of "tree" of possible (in the future) of such algorithms.

It is known that NP-tasks are reduced to NP-complete tasks. The main hope of developers of algorithms is based on the idea that it will be possible to prove the fact coincidence of P and NP classes, it will be possible to solve successfully any tasks of NP class (there are thousands of them now). Word "any" embarrass intuition of the careful observers here. Bright projections sadden two circumstances: firstly, we don't know when long-awaited future will come and, secondly, polynomial "interfusion" of one task with polynomial effective algorithm of other task can give as a result inefficient algorithm. The way out for similar situations is in absence of summation of complexity of algorithm and complexity of procedure of reduction of one discrete task to another. For this purpose it is necessary not to go deep into these traditional bulky procedures, and to develop identical technology of obtaining the answer of discrete special cases of only one extremely universal mass task "Arrangement".

With extremely common conceptual point of view, all without exception in nature of mass and individual math tasks can be considered as a special case of the task "Arrangement". And really the main idea of any task's solution is to ensure to get the combination of the parameters which fully satisfies declared in the input data requirements of the task to its [the necessary exact] answer. It means that in any process of the solution of the any task find optimal arrangement (optimal combination) of all [necessary for solution of this very task] and almost always conflicting properties in a virtual set of information, which is called the "answer to the task".

**Example:** the task «TSP» («Travelling Salesman Problem»). Conditions: it is given graph with $n$ nodes and with any way located forbidden and unforbidden tree edges. The unforbidden edges have any lengths. The purpose of the decision of this task: to find the closed route of the least length which one time passes through all $n$ nodes. Speaking differently, it is necessary to find an optimum combination of $n$ tree edges - optimum in the sense that this combination completely satisfies to all conditions of the task.

## 1.0. Principle «synthesis of the answer of a task»

Strictly speaking, it is possible to synthesize the answer for both discrete and "not discrete" mass task. There is no basic difference here; everything depends on accuracy of algorithm's work and on labor input of this process (on expenses of resources where machine time and memory are the main). The more elements in the declared sets of values of variables, the synthesis is longer, and in "not discrete" tasks such sets have infinite capacity – therefore, process of synthesis will last infinite time. Therefore "synthesis" is rational to apply the principle where other appears less effective – for tasks with discrete sets of independent variables and for NP-tasks.

Inevitable payment for polynomial complexity of synthesis process of the answer which is expressed by quantities, instead of Boolean "yes/no" is a lack of a guarantee that the created answer will always be the exact answer of a solved task.

## 1.1. Environment of a task

The Environment of a task is available to the developer of algorithm virtual model of this task in which step by step, from separate fragments, the required answer is consistently created. Unlike traditional computing approaches it is not so mathematical (analytical, functional) but topological and combinatory model of the system being a task.

One of the integral properties of any unresolved task is uncertainty existing in this task concerning its required answer. This uncertainty, naturally, is always various for different elements of a set "task model" (for different fragments of future answer). Proceeding from physical sense of a concrete special case of the task "Arrangement" uncertainty can be considered from several points of view, i.e. can be traced by several parameters.

Respectively, it will be possible to define many different parameters expressed computationally for each set of values of parameters. They can be interpreted as elementary colors of variable brightness. These "colors" can be reduced

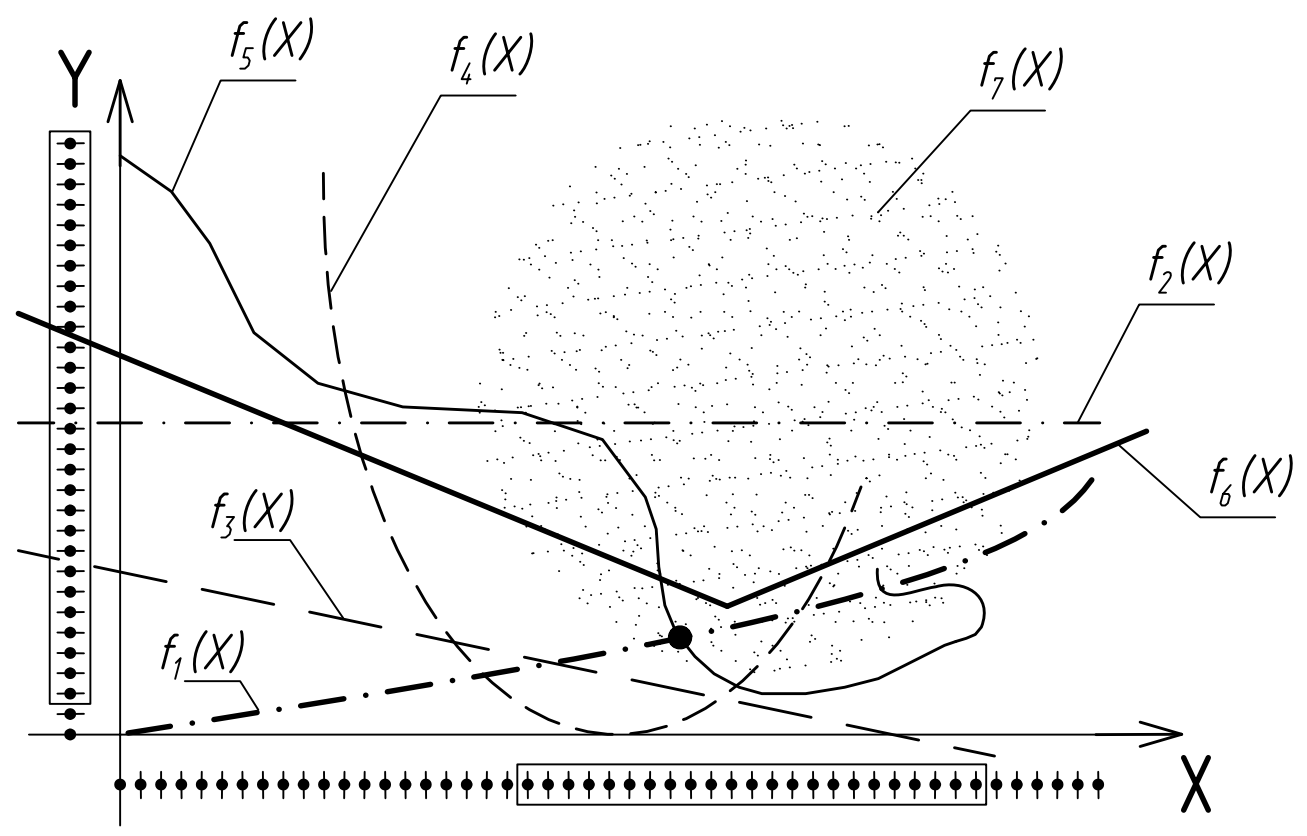

*Here:*

X and Y – *independent and dependent variable accordingly*

♦♦♦♦♦♦♦♦ – *set of variable's values*

▭ – *range of definition of variable, declared by the user of algorithm*

$f_1(X) \ldots f_7(X)$ – *functional dependencies, declared by the developer and concretized by the user*

● – *exact answer of this task*

*Fig.1 Graphic interpretation of the general principle of detection of the exact answer on an example of certain task with two continuous variables (the classical analitical-computing approach)*

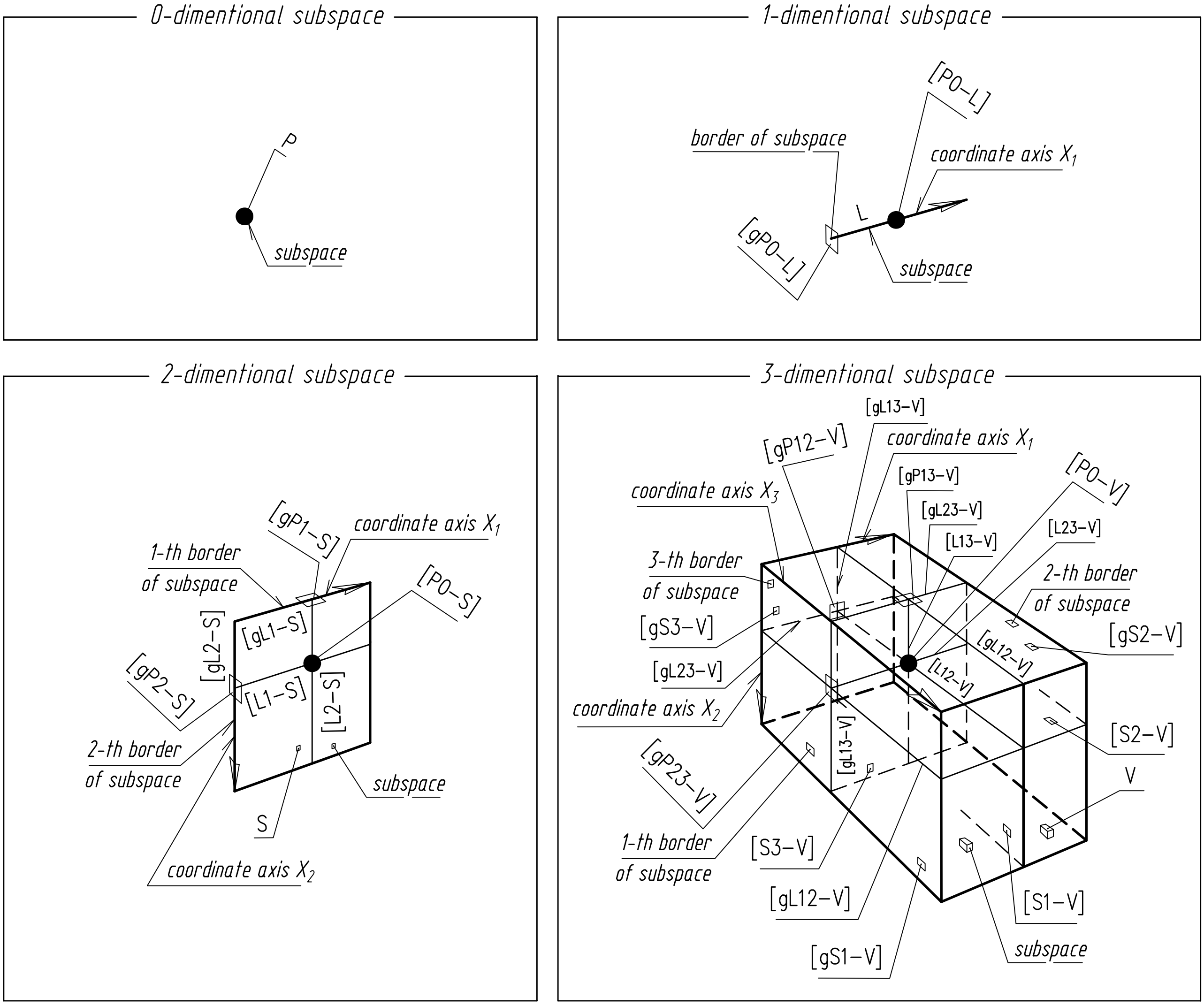

*Here:*

$X_1$, $X_2$, $X_3$, ... – *variables which are considered in a task (or, that the same, coordinate axes in the Environment)*

### elements of 0-dimentional subspace

P – *point*

### elements of 1-dimentional subspace

L – *line*
[P0–L] – *point-0 (the point belongs to line L)*
[gP0–L] – *border point-0 (the point belongs to line L)*

### elements of 2-dimentional subspace

S – *plane*
[L1–S] – *line-1 (the line is parallel $X_1$)*
[L2–S] – *line-2 (the line is parallel $X_2$)*
[P0–S] – *point-0 (the point belongs to plane S)*
[P1–S] – *point-1 (the point belongs to line-1)*
[P2–S] – *point-2 (the point belongs to line-2)*
[gL1–S] – *border line-1 (the line coincides with $X_1$)*
[gL2–S] – *border line-2 (the line coincides with $X_2$)*
[gP1–S] – *border point-1 (the point belongs to the border line-1)*
[gP2–S] – *border point-2 (the point belongs to the border line-2)*

### elements of 3-dimentional subspace

V – *volume*
[S1–V] – *plane-1 (the plane is perpendicular $X_1$)*
[S2–V] – *plane-2 (the plane is perpendicular $X_2$)*
[S3–V] – *plane-3 (the plane is perpendicular $X_3$)*
[L12–V] – *line-12 (the line is perpendicular $X_1$ and $X_2$)*
[L13–V] – *line-13 (the line is perpendicular $X_1$ and $X_3$)*
[L23–V] – *line-23 (the line is perpendicular $X_2$ and $X_3$)*
[P0–V] – *point-0 (the point belongs to volume V)*
[P1–V] – *point-1 (the point belongs to plane-1)*
[P2–V] – *point-2 (the point belongs to plane-2)*
[P3–V] – *point-3 (the point belongs to plane-3)*
[P12–V] – *point-12 (the point belongs to line-12)*
[P13–V] – *point-13 (the point belongs to line-13)*
[P23–V] – *point-23 (the point belongs to line-23)*
[gS1–V] – *border plane-1 (the plane is perpendicular $X_1$)*
[gS2–V] – *border plane-2 (the plane is perpendicular $X_2$)*
[gS3–V] – *border plane-3 (the plane is perpendicular $X_3$)*
[gL12–V] – *border line-12 (the line is perpendicular $X_1$ and $X_2$)*
[gL13–V] – *border line-13 (the line is perpendicular $X_1$ and $X_3$)*
[gL23–V] – *border line-23 (the line is perpendicular $X_2$ and $X_3$)*
[gP1–V] – *border point-1 (the point belongs to the border plane-1)*
[gP2–V] – *border point-2 (the point belongs to the border plane-2)*
[gP3–V] – *border point-3 (the point belongs to the border plane-3)*
[gP12–V] – *border point-12 (the point belongs to the border plane-3 and line-12)*
[gP13–V] – *border point-13 (the point belongs to the border plane-2 and line-13)*
[gP23–V] – *border point-23 (the point belongs to the border plane-1 and line-23)*

*Рис. 2 Geometrical interpretations of subspaces by means of which it is possible to create the Environment of any task on discrete structures (are used the cartesian coordinates)*

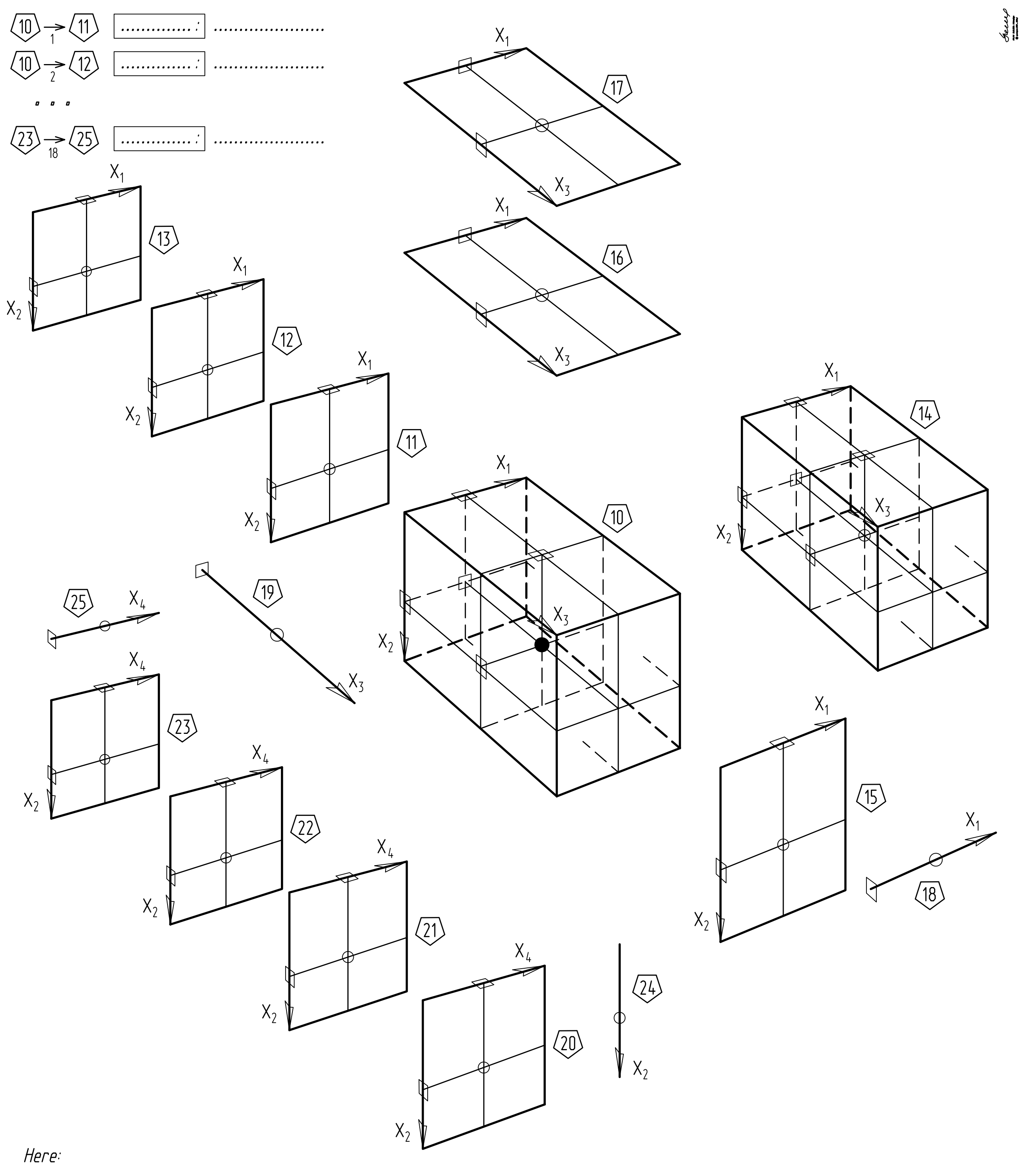

*Here:*

*$X_1$, ... $X_4$ - variables of this task (it is necessary to create an optimum combination of their values from point of view of the declared limitations)*

*(10) - main subspace (step-by-step synthesis of as more as possible successful and balanced answer of a task here is made)*

*(11) - Potential of main subspace (i.e. the starting position of all Marks of the main subspace)*

*(12) - Counterpotential of main subspace (i.e. the starting position of all algorithmic Interdictions of the main subspace)*

*(13)...(19) - Factors 1, 2, 3, 4, 5, 6 and 7 of main subspace*

*(20) - 1-th additional subspace*

*(21) - Potential of 1-th additional subspace*

*(22) - Counterpotential of 1-th additional subspace*

*(23)...(25) - Factors 1, 2 and 3 of 1-th additional subspace*

*(A) →c (B) - Contact dependence of B from A (i.e. all those changes in B which appear after the changes accomplished by algorithm in A)*

*Рис. 3 Geometrical interpretation in the cartesian coordinates of the Environment of a certain NP-complete task with 4 variables*

(by means of certain weight functions, systems of competitions, etc.) to one numerical size, to a certain integrated parameter (to "resultant color") which, in relation to idea of metaedroalgorithm, describes all points of [discrete] model of a task on discrete structures. Then a randomness (irregularity) of unequal values of parameters of points of this model from depressing inconvenience becomes unexpected invaluable advantage: in general case all possible fragments of the created answer at every moment of work of algorithm have different quantitative values of "resultant color". Consequently, among them it will be possible to select those fragments which will be more pertinent and useful in future answer of a task.

Consequently, it is possible to use uncertainty as instrument for the solution of a task in the course of gradual reduction and elimination of uncertainty existing in its answer.

The problem of the developer of algorithm is to define "resultant color" both which its shades and sizes of intensity will be pertinent and useful in this step by step created to "the best combination of properties". Than actual working ideas of the developer will be more adequate and than their physical realization is more successful, especially a version of metaedroalgorithm will be effective.

The minimal set of concepts which is required at construction of the Environment of any concrete mass task on discrete structures:

**Definition 1**: N-dimensional space of a task - the Cartesian product of $n$ sets, where $n$ <= N. Each of these N sets is the set of all the values of one of the N variables that exist in the task. The value of $n$ is the quantity of the parameters that completely define required answer of the task (i.e. $n$ is the quantity of unknowns in the task).

**Definition 2:** senior space [of a task] - the Cartesian product of $n$ sets where $n$ there is the quantity of unknowns in this task.

**Example:** In the task of "TSP" the senior space is the set consisting of $n^n$ of all possible routes from the point of view of combination theory (each such route is formed of $n$ tree edges).

**Definition 3:** subspace [of a task] - any ordered subset of that set which is N-dimensional space of the given task.

**Definition 4:** junior space [of a task] - the ordered set of values of all independent variables and all dependences between them, which consider in a task.

**Example:** set of all possible from the point of view of combination theory of the graph node couples in the task "TSP" whose conditions are set: lengths of tree edges, existence of the forbidden edges, various additional conditions.

The junior space is one of the most important for the developer of algorithm of subspaces of a considered task: the user of algorithm can deal only with junior space.

**Definition 5:** Environment - a virtual model of junior space of a task on the discrete structures, which consist of a finite set of subspaces influencing at each other.

**Definition 6:** Relation - the one concrete dependence of any nature (quantitative, functional, logical, topological, combinatory etc) which is considered as existing in the Environment between a finite set of values of variables considered in the Environment.

**Examples** of the Relations in task "TSP": length of an tree edge (quantitative Relation); "the tree edge can connect only two graph nodes" (logical Relation); relative positioning of not forbidden and forbidden tree edges (topological Relation); belonging of these or those edges to these or those subsets of edges (combinatory Relation).

**Definition 7:** Factor - value of one parameter, appropriated to one concrete value of one variable.

**Example** of the quantitative Factor in the task "TSP": length of an tree edge.

**Definition 8:** Summary decision - the required answer of a task which is created (is synthesized) by algorithm.

**Example:** one finite set of tree edges in the task "traveling salesman problem", being the required closed route.

**Definition 9:** Single decision - the smallest of fragments of any formally possible Summary decision of a task making physical sense (or, it is the same, any of points of junior space).

**Example:** one separately taken tree edge in the task "TSP".

**Definition 10:** Transition point - the point of the Environment declared by algorithm as the next Single decision.

Clearly that its answers are exactly in junior space of any task "in the look sorted on fragments" without an exception. All these answers "in assembled form" make its [discrete] N-dimensional space, but it is not possible to find something certain in it (before the positive solution of the dilemma «P ? NP»). Junior space is different, which by definition is always a polynomial set (since it always manually is set by the user of algorithm) and can be many times seen by algorithm quickly.

**Definition 11:** Potential point - a point of the Environment which can (but optional has to) become the Single decision in the created answer of a task.

**Example:** any not forbidden (by the user of algorithm or by algorithm) tree edge in the task "TSP" which isn't already appointed Single decision.

**Definition 12:** Forbidden point - a point of Environment which can not or/and for any reasons should not become a Single decision in the created answer of a task.

**Example:** any forbidden (by the user of algorithm or by algorithm) tree edge in the task "TSP".

**Definition 13:** Marker - a designation in subspace of the fact of existence of something: designation of Relations, of Single decision in the Summary decision etc.

**Definition 14:** Interdiction - a designation in subspace of the fact of prohibition of existence of something.

**Definition 15:** Step - the set of all operations made by algorithm in an interval between two next states of the Environment (between two next changes of the Environment: change of the Environment is the announcement algorithm of the next Point of transition).

Fig. 2 and 3 illustrate a basic way of creation of subspaces and Environment. If necessary it is possible to apply any other systems of coordinates, the bigger quantity much more precisely and subspaces influencing at each other and, respectively, more features of a task which are fully considered in the course of its decision. There is only one restriction here: the more parameters, values of these parameters and influences of these values at each other, the objectively less chances to make the answer with a desirable combination of values of parameters are possible.

## 1.2. Changes of Environment

At the end of each Step the metaalgorithm appoints the next Point of transition, i.e. the next Single decision. This "independent" change causes a final set of "dependent" changes and this set is possible to designate further as "Environment reaction". Each such dependent change is carried out by means of finite set of «Contact dependences» which is stipulated by the developer of algorithm.

**Definition 16:** Contact dependence [of subspace B from subspace A] - finite set of those or other actions made by algorithm in subspace B depending on those or other changes, which occurred in subspace A.

The elementary form of Contact dependences is any calculated functional dependence. But much more capable, flexible and universal are those actions which the computer is capable to make is a creation, examination, sorting and change of any final sets, comparison of values of parameters, logic constructions etc. And also, certainly, any calculations and use of any functional dependences.

However, for the successful solution of a task only the "completeness" and "efficiency" parameters of the created answer are important.

**Definition 17:** completeness of the Summary decision - a number from an interval (0, 1), which is the relation of quantity of all Single decisions of a task received by algorithm to quantity of Single decisions in [objectively existing] exact answer of the same task.

**Definition 18:** efficiency of the Summary decision - a number from an interval (0, 1) growing out of comparison of the Summary decision received by algorithm and the exact answer of this task without the "completeness of the Summary decision" parameter.

The quantity and proximity to each other of approximate answers of NP-class tasks always allow to hope that the metaedroalgorithm, which has been successfully worked by the more or less reasonably and adequately acting developer, will synthesize such answers which more than satisfy even the most distinguished requirements of practice.

Any business which doesn't manage to be carried out at once, can be made step by step, in the form of a finite great set of simpler and therefore more feasible actions. It can be said of artificial creation of the answer of a discrete task "Arrangement", this answer always consists of a finite set of fragments (from the Single decisions included in this answer). There is always a full set of all Single decisions, possible in a solved task at the disposal of the developer of algorithm (i.e. in the Environment).

Step-by-step synthesis of the answer of a task on discrete structures (step-by-step change of a state of the Environment) is completely similar to movement through an unfamiliar labyrinth. It can't essentially be crossed at one stroke, it is necessary to stop on each met fork and to choose one of many offered ways. Moreover, Environment represents such confusion of the resolved and forbidden ways which is not only changes the subsequent chances of its successful overcoming depending on the undertaken actions on each of forks as it happens in any other virtual or physical labyrinth. Environment also changes quantity and a relative positioning of the remained passes and forks.

Overcoming a labyrinth (such as Environment) is very difficult problem. For successful synthesis of the whole fragments, each of which influences on other fragment, and each combination of such fragments influences each other combination of these fragments, the algorithm has to own the most important of an exponential set of these influences and dependences. Any algorithm will face here the fundamental information phenomenon, with certain viscous and faceless "the entropy resistance" to all what will try to make here. It is inevitable payment for possibility of scanning of the polynomial Environment instead of exhaustive search of exponential N-dimensional space of the same task.

**Definition 19:** Latent risk - a possibility of transformation into the Forbidden point of at least one of those points of the main subspace of the Environment which are unique (i.e. not replaceable other points) Single decisions of the created Summary decision.

If the metaedroalgorithm at least once didn't manage to avoid transformation into the Forbidden point of the unique Single decision of future Summary decision, the completeness parameter of the created Summary decision appears less than 1 (i.e. the received answer becomes unsuitable for practical use).

## 1.3. Navigation nucleus of metaedroalgorithm

For the successful solution of a task it isn't enough the only adequate model, i.e. successfully made Environment in which the necessary and sufficient set of its reactions (changes) is provided at influence from the outside. The algorithm has to be able to appoint each time the most successful (from the point of view of its influence on already known and yet not appointed Single decisions) the next Point of transition.

What it has to be guided on so that consequences of each of m of such elections wouldn't bring to absolutely unsuitable or to less successful Summary decision? How to make it if the required Summary decision and its parameters by which it would be possible to be guided beforehand, hasn't exist yet? The metaedroalgorithm has to strive in every possible way for perfection (i.e. for the exact answer of a solved task) about which there can be no information. In a task there is an exponential quantity of unknown applicants distributed in space for perfection. The detached onlooker (i.e. the person, "the one who solves a task") in general case knows and learns nothing about the exact answer of a solved task.

The huge problem that at any announcements of Points of transition Potential points in the Environment will disappear much earlier, than the Summary decision will be created. Because at successive of any actions causative dependences, big sets of uncontrollable parameters, opportunities, situations, ways, interactions, "labyrinth forks" are the main.

In other words, algorithm needs the certain means providing obtaining of objective information about a situation in Environment where they are now and, secondly, something that carries out adoption of as much as possible successful decision what to do farther in the conditions of existing in this environment. The first of this obligatory set for definiteness it is possible to call the "sensors", the second – "a navigation nucleus of metaedroalgorithm".

There are bases to assume that in any chaos there are steady regularities, inevitably showing "patterns". But these regularities are much less obvious and more capricious, than "calculated" (analytical, functional) dependences. One of the basic tools, allowing to be guided in exponential sets of combinatory and cause and effect dependences, can not be functions and other pretty rigid correlations, and certain universal and at the same time quite utilitarian paradigms.

**Definition 20:** paradigm [of approach "synthesis of answers of tasks"] - the general principle by means of which it is possible to reveal degree of points suitability of task space to be the required answer of a task.

**Definition 21:** valency - dependence between the Environment point (i.e. Single decision) and the closest to the exact answer of a task Summary decision; is expression of desirability degree of Single decision existence as a fragment of this Summary decision: the more desirability of such existence, the more quantitative expression of valency.

As each point of Environment has valency, this property is possible to consider as the certain function determined on the set «points of Environment».

There are as much such paradigms, how many happens in the nature of spectrum's colours, notes, the main smells, tastes, etc. – seven. Three of them are rather obvious: last experience, forecasting of the future and ban on commission of inadmissible actions in the present. But true and driving, capable in current state of the Environment to foresee the most possible remote consequences of these or those changes are other fundamental paradigms. There are else as minimum four such paradigms, but the volume of this article doesn't allow to go deep into these questions.

## 2. Examples of adaptation of basic metaedroalgorithm to specific tasks on discrete structures

### 2.0. The task «Heap»

Conditions of this task: there is a some finite set of subjects of any weight. It is necessary to divide this set into two subsets of equal weight.

In the Environment of the given task (see Fig.4) $X_1$ is the variable "subjects in first half of heap", $X_2$ is variable "subjects in second half of heap", $X_3$ is variable "weight of a subject". At start of algorithm in subspace 10 there are no Markers.

<u>Contact dependence of subspace 11 from subspace 10</u>: if the point in subspace 10 appears the Point of transition, then algorithm moves the Marker from subspace 11 into the same point in subspace 10.

<u>Contact dependence of subspace 20 from subspace 10</u>: if the point in subspace 10 appears the Point of transition, then algorithm moves the Marker from subspace 20 into the appropriate point in subspace 21.

<u>Contact dependence of subspace 22 from subspace 10</u>: if the point in subspace 10 appears the Point of transition, then algorithm moves the Interdiction from subspace 22 into the same point in subspace 20.

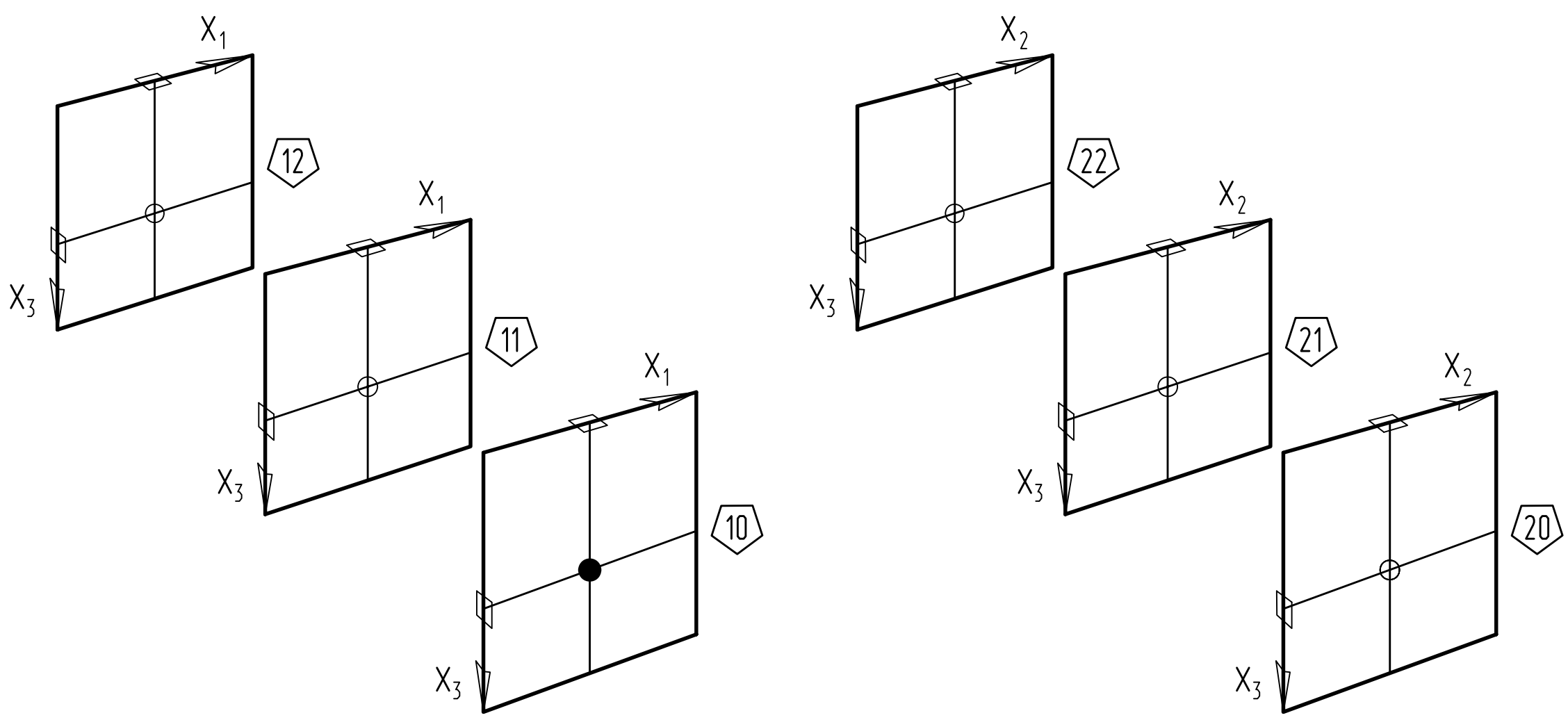

*Here:*

- $X_1$ - *variable "subjects in first half of the heap"*
- $X_2$ - *variable "subjects in second half of the heap"*
- $X_3$ - *variable "weight of a subject "*
- (10) - *main subspace (the fact of accommodation of a subject in the first half of heap here is registered)*
- (11) - *Potential of main subspace*
- (12) - *Counterpotential of main subspace*
- (20) - *additional subspace (the fact of accommodation of a subject in the second half of heap here is registered)*
- (21) - *Potential of additional subspace*
- (22) - *Counterpotential of additional subspace*

*Fig. 4 The Environment of the some individual case of the task "Heap"*

⟨10⟩ →(1) ⟨11⟩ | at node appears 1-th or 2-th tree edge: | Marker from ⟨11⟩ moves into ⟨10⟩

⟨10⟩ →(2) ⟨12⟩ | at node appears 1-th or 2-th tree edge: | that Interdiction from ⟨12⟩, which corresponds to the Marker mentioned in ⟨10⟩ →(1) ⟨11⟩, moves off ⟨12⟩

⟨10⟩ →(3) ⟨12⟩ | at node appears 2-th tree edge: | all that Interdictions from ⟨12⟩, which can remove from participation in game all other tree edges at this node, moves into ⟨10⟩

⟨10⟩ →(4) ⟨11⟩ | at node appears 2-th tree edge: | all that Markers from ⟨11⟩, which corresponds to the Interdictions mentioned in ⟨10⟩ →(3) ⟨12⟩, moves off ⟨11⟩

## set M "independent variables"

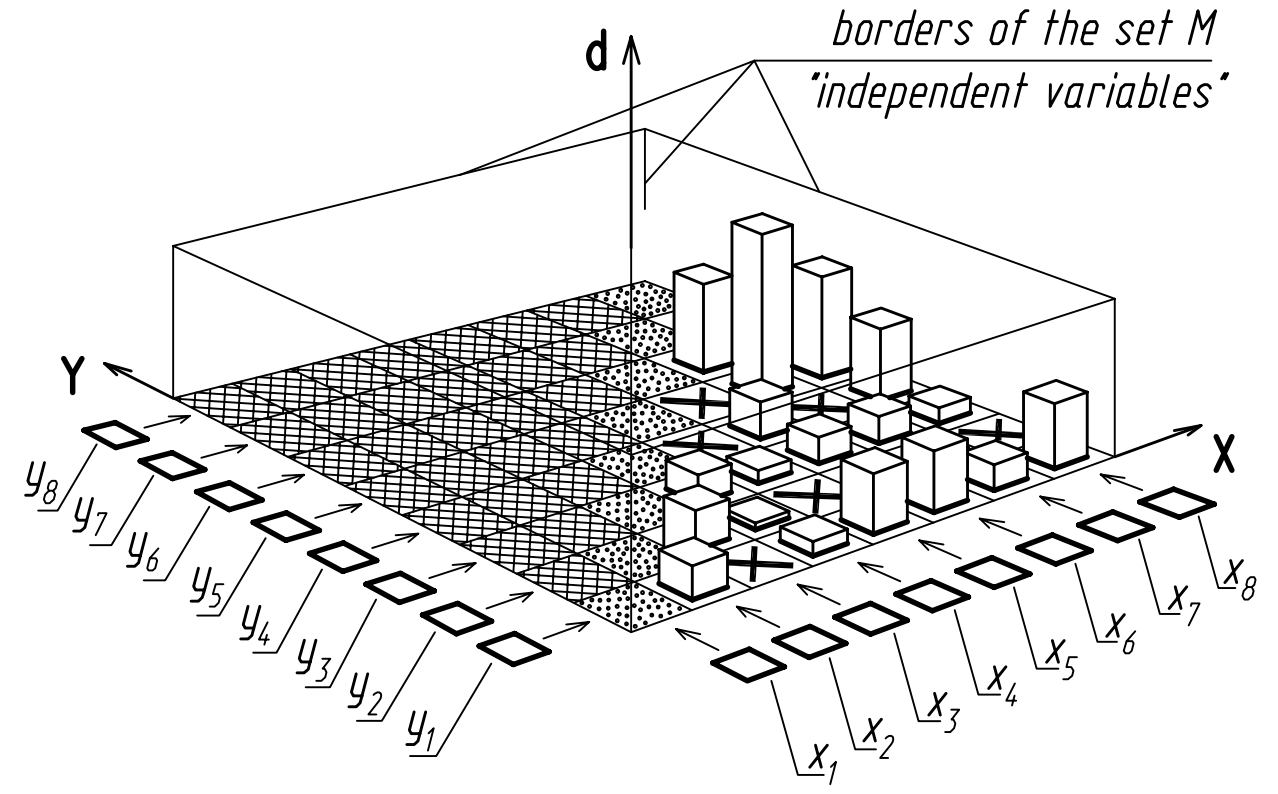

## Environment

(a state for a moment "the end of the interval 1 of the Step 1")

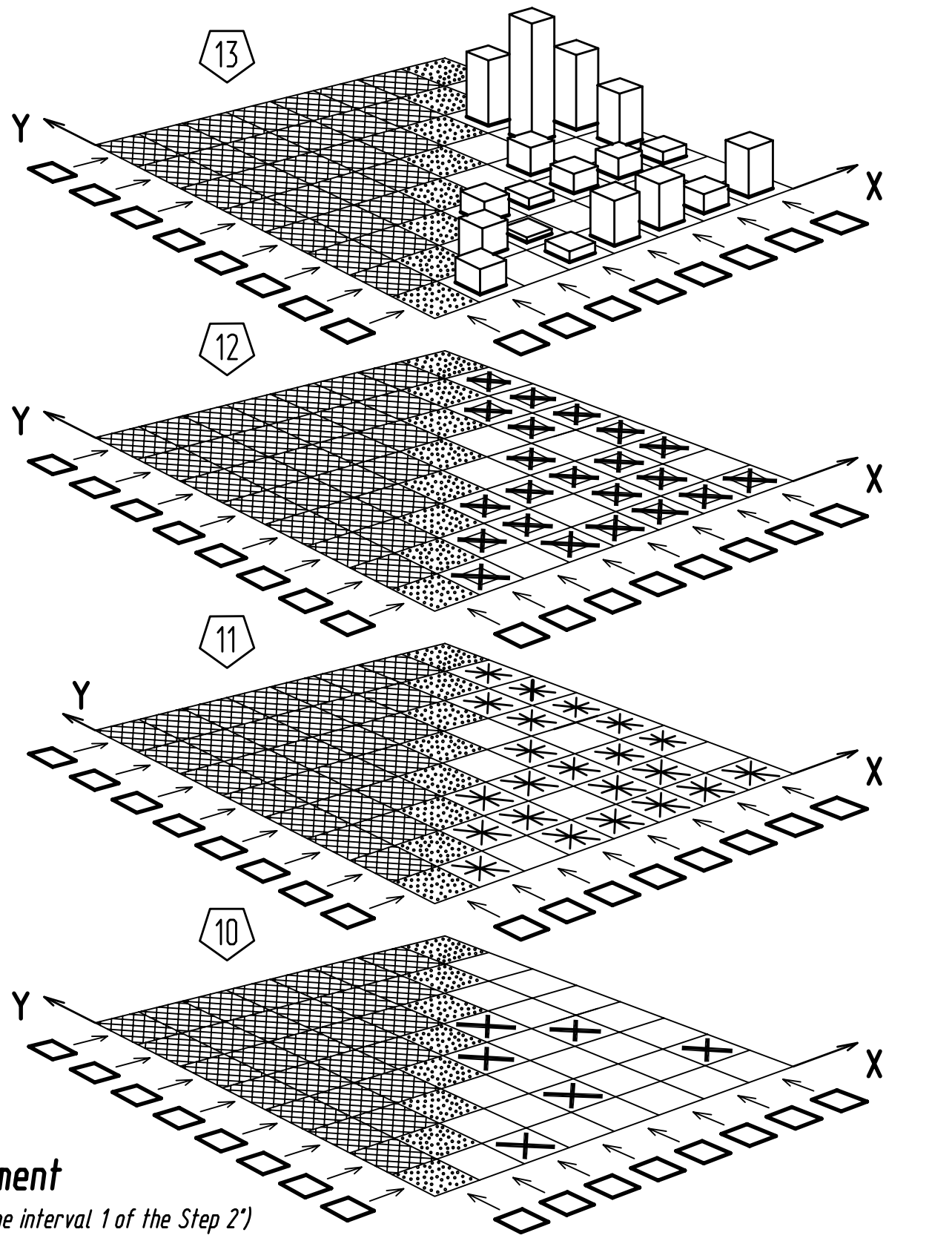

## Environment

(a state for a moment "the end of the interval 1 of the Step 2")

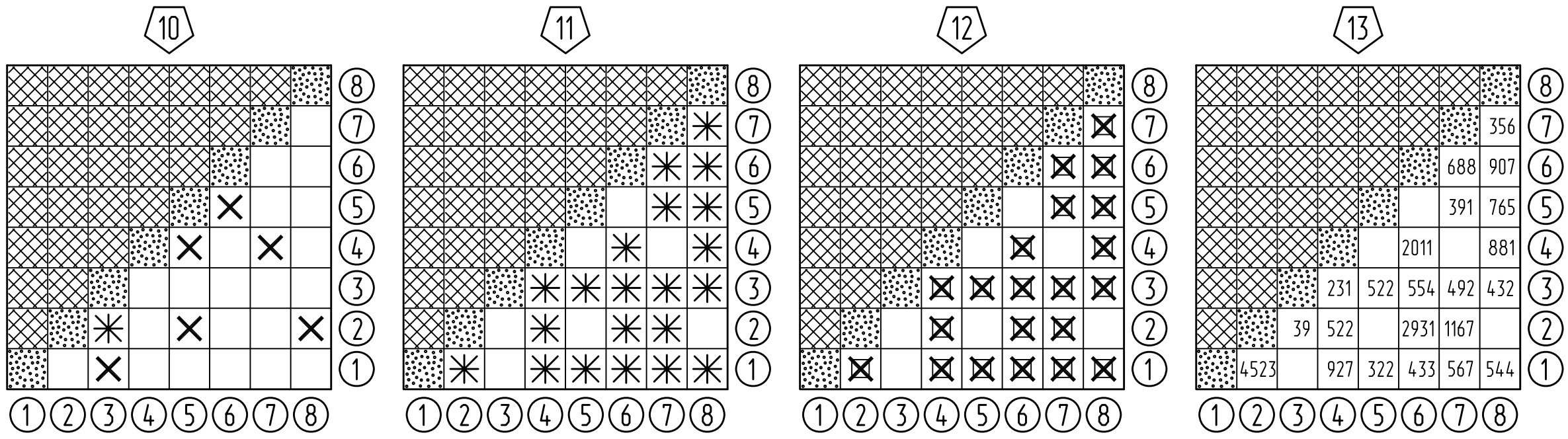

*Here:*

$x_i$ - i-th member of set "independent variable X"; parameter $P_1$ appropriated to this member: "i-th initial graph node"

$y_j$ - j-th member of set "independent variable Y"; parameter $P_1$ appropriated to this member: "j-th final graph node"

$d_{ij}$ - ij-th member of set "independent variable d"; parameter $P_1$ appropriated to this member: "length of a tree edge between i-th initial and j-th final graph nodes"; members $\{x_i, y_j, d_{ij}\}$ form subset E "unforbidden tree edges"

- ij-th member of set T "tree edges which are forbidden by the user [in the given individual task "TSP"]" (an user Interdiction)

- ij-th member of set T "tree edges which are forbidden by the algorithm [in the given individual task "TSP"]" (an algorithmic Interdiction)

- nonexistent tree edge (member of set UE)

- set member of set UL of not making sense tree edges

- Point of transition [to the following state of Environment] (i.e. the next Single decision)

⟨10⟩ - main subspace

⟨11⟩ - Potential of main subspace

⟨12⟩ - Counterpotential of main subspace

⟨13⟩ - Factor 1 (i.e. lengths of tree edges $d_{ij}$)

⟨A⟩ →(C) ⟨B⟩ - Contact dependence of B from A (i.e. all those changes in B which appear after the changes accomplished by algorithm in A)

Рис. 5 The Environment of the some individual case of the task "TSP" (n=8)

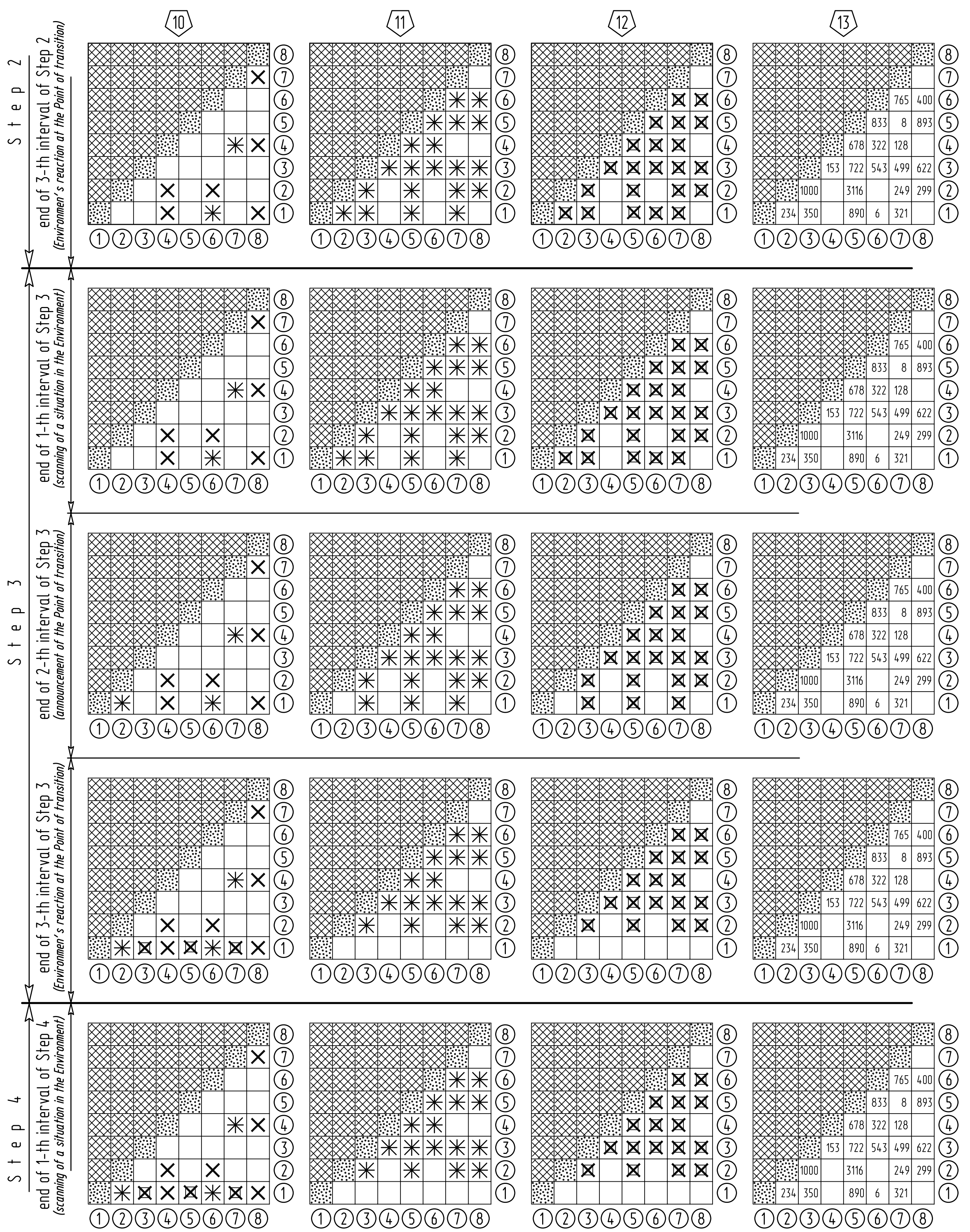

*Here:*

○ – *marking of the graph node*

– *accordingly: forbidden by the user, nonexistent and not making sense tree edge*

$d_{ij}$ – *value of parameter declared by the user of algorithm (length of the tree edge between the specified graph nodes)*

– *Point of transition [to the following state of Environment] (i.e. the Single decision)*

– *algorithmic Interdiction*

*Fig. 6 Changes of the Environment of an individual case of the task "TSP" (n=8)*

Other versions of subspaces, additional subspaces can be applied to the accounting of thinner dependences, requirements and limitations, and also other Contact dependences.

On each Step the navigation nucleus of metaedroalgorithm investigates the current situation in all subspaces of the Environment, formalizes it in a quantitative form and reveals the current provision of a maximum of the valency function changing on a set of points of the main subspace. The potential point with the greatest value of this function is the Point of transition of this Step (in order to avoid unnecessary ideological complication of metaedroalgorithm and emergence of bulky branchings, on each Step only one Point of transition appears).

Such standard Steps repeat until the difference between the sums of weight s of subjects in both heaps decrease. After that, because of ideological simplicity of a task, a polishings technology can be useful i.e. transfer of several subjects suitable on weight in "others" heap for even bigger reduction of this difference.

Unlike of many other NP-tasks in the task "Heap" very easily the fact of existence of the exact answer can be defined. Quantitative parameter of the exact answer is a half of total weight of all subjects and, therefore, degree of perfection of the Summary decision created by algorithm (the "efficiency" parameter, i.e. discrepancy degree decides on the exact answer). It is very convenient for working off in these or those versions of metaedroalgorithm of new ideas concerning level of accuracy of its work.

The Summary decision will always be the suitable answer for use here. The "completeness" parameter in this task doesn't make sense and the quantitative expression. For working off of new ideas concerning level of completeness of answers synthesized by metaedroalgorithm, is more suitable other "a fly the drosophila" of computer sciences, task "Travelling Salesman Problem".

## 2.1. The task «Travelling Salesman Problem»

In the Environment of the given task (see Fig.5) $X_1$ is the variable "initial node of an tree edge", $X_2$ is variable "final node of an tree edge", $X_3$ is variable "length of an tree edge". Before start the user of algorithm appoints the forbidden tree edges and lengths of the not forbidden edges.

Other processing methods in algorithm essentially don't differ from similar receptions in algorithm for the task "Heap" since they are two versions of the same algorithm of the solution of a task "Arrangement". Simply in one case two as much as possible successful sets are synthesized (from the point of view of equality of weights of this sets), in other – one set of tree edges forming a shorter closed route.

× × ×